# *General-purpose and dedicated regimes in the use of telescopes*

Jérôme Lamy (LISST, Université de Toulouse, CNRS, France )
Emmanuel Davoust (LATT, Université de Toulouse, CNRS, France)

**Abstract :** We propose a sociohistorical framework for better understanding the evolution in the use of telescopes. We define two regimes of use : a general-purpose (or survey) one, where the telescope governs research, and a dedicated one, in which the telescope is tailored to a specific project which includes a network of other tools. This conceptual framework is first applied to the history of the 80-cm telescope of Toulouse Observatory, which is initially anchored in a general-purpose regime linked to astrometry. After a transition in the 1930s, it is integrated in a dedicated regime centered on astrophysics.  This evolution is compared to that of a very similar instrument, the 80-cm telescope of Marseille Observatory, which converts early on to the dedicated regime with the Fabry-Perot interferometer around 1910, and, after a period of idleness, is again used in the survey mode after WWII. To further validate our new concept, we apply it to the telescopes of Washburn Observatory, of Dominion Astrophysical Observatory and of Meudon Observatory. The uses of the different telescopes illustrate various combinations of the two regimes, which can be successive, simultaneous or alternating. This conceptual framework is likely to be applicable to other fields of pure and applied science.

**Keywords :** telescope, general-purpose regime, dedicated regime, practices

## I. Introduction

The role played by the telescope in the evolution of our knowledge of astronomy often remains hidden, even though it is obviously essential.

A certain number of studies have been carried out about astronomers' light collectors, like William Hershel's telescope[1], for example, or the 12-inch telescope at Lowell Observatory that was used to search for transuranian planets and led to the discovery of Pluto.[2] More contemporary instruments have also been the subject of historical analysis. Brad Gibson traced the technical difficulties linked to the construction of the Canadian liquid mirror telescope.[3] The history of the British Isaac Newton telescope, in contrast with that of more recent British telescopes in the Canary Islands, reveals the influence, in both periods, of the scientific and technical context of such a project.[4] As for the itinerary that led to the construction of the Anglo-Australian 150-inch telescope, it reflects the political stakes which the United Kingdom had to face at the time, and in particular the difficult choice between two possible partners.[5] Trudy E. Bell's study of the construction of large telescopes by the Warner & Swasey Company provides evidence for the ties between science and industry in the field of astronomical instrumentation.[6]

In France, Philippe Véron has shown the difficult installation of the equatorial in the eastern tower of Paris Observatory.[7] Emmanuel Davoust has described the construction and use of several telescopes at Pic du Midi Observatory in a very special local environment that required unusual human qualities.[8] The work of Audouin Dollfus on the large telescope at Meudon shows the successive uses of an instrument of exceptional dimensions.[9] William Tobin has related the history of the Foucault telescope at Marseille Observatory and has tried to learn its lessons for the management of future astronomical projects.[10]  The large number of these analyses does not, nonetheless, exhaust the interest of historical research into a specific instrument.

The goal of the present paper is to examine how telescopes structure, or otherwise influence, research in astronomy, and to establish a general pattern for their often changing role in assisting (or leading) astronomical research. We are concerned with telescopes that were considered large at the time of their first light. The telescopes selected for study are all reflectors, except the Meudon telescope which is a refractor.

The historical period of interest starts with the beginning of astronomical photography (the 1880s), and ends with the advent of modern large telescopes (the 1960s or 70s).

We first describe a general conceptual framework for evaluating the role of telescopes, and define two regimes, the regime of general use (or survey regime), and the dedicated regime. We then make a detailed assessment of the role of one specific instrument, the 80-cm telescope of Toulouse Observatory, and establish its successive roles. We then move on to a comparison with another telescope of the same size in a very similar environment, revealing similar roles, but with a different chronological order. Finally, the pattern which emerges from this comparison is tested on the history of other telescopes, taken from the literature, thus providing a conceptual framework for analysing the history of astronomical instruments.

## II. The general and dedicated regimes of use

The large reflectors and refractors built toward the end of the 19$^{th}$ century by and large match the technical and scientific criteria set up by astronomers. In this sense, large telescopes are comparable to the generic instruments defined by Terry Shinn[11] : their flexibility, optical quality and their ease of use allow them to satisfy a large range of demands. Telescopes thus fit in a specific technological context, one strongly marked by the determination of the position of celestial objects -- astrometry. This first regime of use, which we call "general regime" or "survey regime", considers observing as a global activity of the observatory. The instruments are not dedicated to a very specific task, they fit in a general scientific policy as defined (explicitly or implicitly) by the director, that of gathering measurements. In a sense, it is a survey mode, where the instruments are put to the task of measuring whatever can be measured with them. One can thus state that it is the instrument, not the astronomer, who drives the science that is made. The astronomer is content to observe using the available tools.

An example of this general regime can be found in the annual report of Toulouse Observatory for 1885-86, which is organized around the instruments (see Fig. 1). There are four telescopes, an astronomer is in charge of each of them, and they all participate to the best of their performances in the observation of a series of targets. Except for the 33cm telescope, they are all used for several programs.

At the other extreme is the dedicated regime, which corresponds to very focused and more oriented scientific practices. Here again, the technical, scientific and political context shape and inflect the uses of the telescope. A specific kind of celestial target, a new dynamic recruit in the team of astronomers, the emergence of an innovative technology (e.g. the microphotometer), of a new scientific program or of an entirely new field which compels teams or institutes to reorient their activities, are as many factors that can impose a dedicated regime for certain instruments. The telescope is then associated with a definite practice, its mechanical and optical performances are overhauled, a new auxiliary instrumentation is acquired in response to new observational goals. The general scientific policy of the observatory is no longer the dominating factor in the use of the telescope; it is rather a combination of new performances of the telescope and its tailoring to a specific program.

An example of this second regime is found in the annual report of Toulouse Observatory for 1957-58, which is now organized by scientific field (see Fig. 2). That of astrophysics is divided into two programs, which makes use of three telescopes, only one of which is local. The 60cm telescope is located at Pic du Midi Observatory and the 120cm on at Haute-Provence Observatory. In other words, the astronomers do not satisfy themselves with the locally available telescopes to pursue their scientific program. The latter clearly drives the use of the telescopes.

As pointed out by a referee, the dichotomy in the roles of telescopes can be seen in different ways. This can be the tension between astronomers who want to undertake research on a specific topic of their choice and those who use the available instruments in the purpose for which they were originally designed. This can also be the distinction between scientific programs shaping the instruments and the available instruments determining the programs. But, from a sociohistorian's point of view, it is essential to consider the telescope as an inanimate actor with a role to play in the conduct of scientific research.

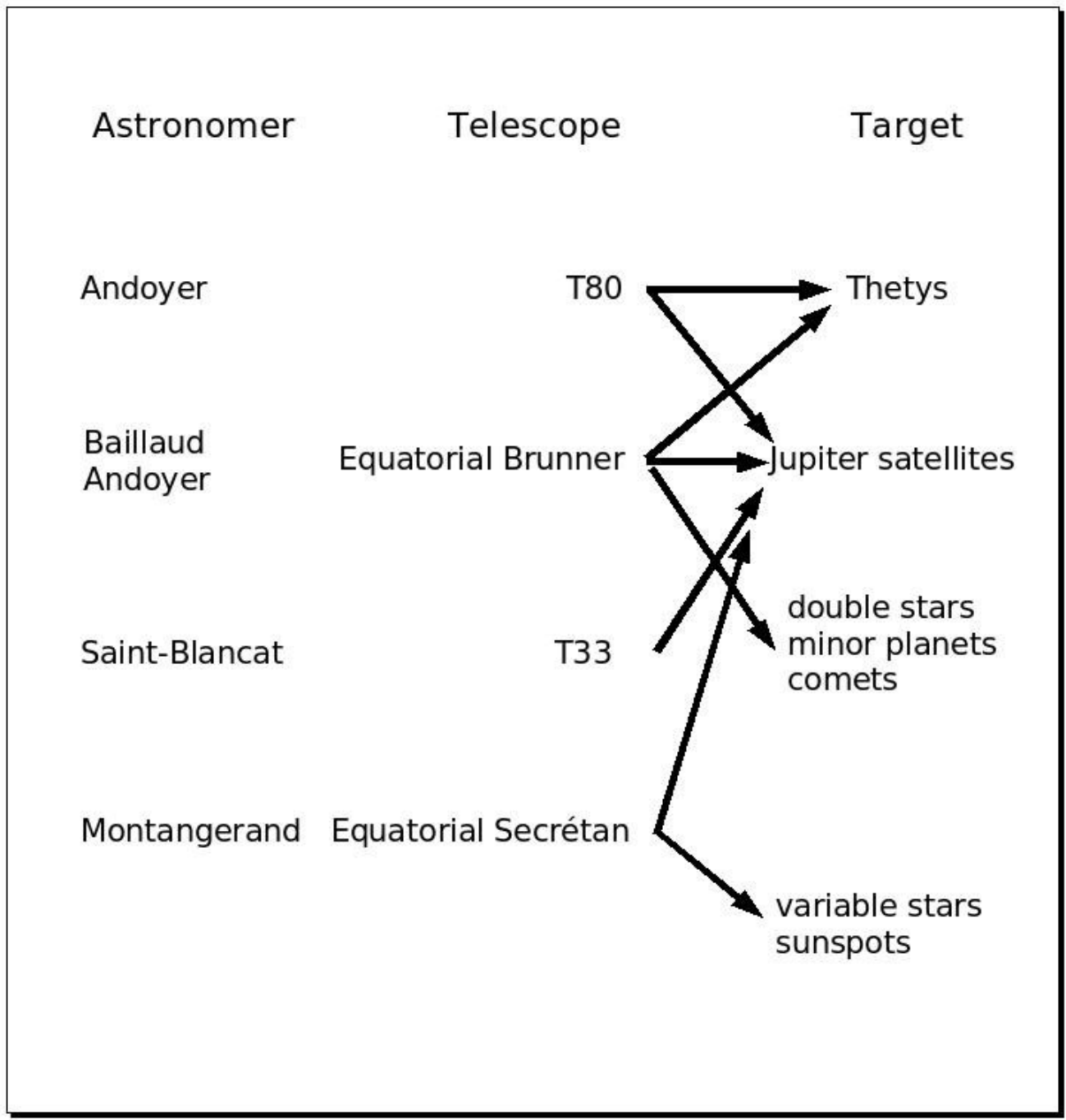

*Fig. 1 : The general-purpose regime at Toulouse Observatory in 1886-87*

At any rate, we do not propose these two, generalist and dedicated, regimes as two rigid ideals, to which all telescopes have to conform. The main point of this paper is to show the flexibility of these concepts, allowing them to shed a new light on the most diverse situations. One of the characteristics of these two regimes of use is precisely the diversity of practical situations and thus of possible combinations of the two regimes : successive, simultaneous, alternating. We insist on the point – and the historical examples that we present bear this out – that there is no unique historical process setting once and for all the uses of a telescope in a quasi-teleological order, moving it in time from the general to the dedicated regime. Again, our empirical approach accounts for the large variety of historical situations, while offering a coherent framework for their sociohistorical understanding.

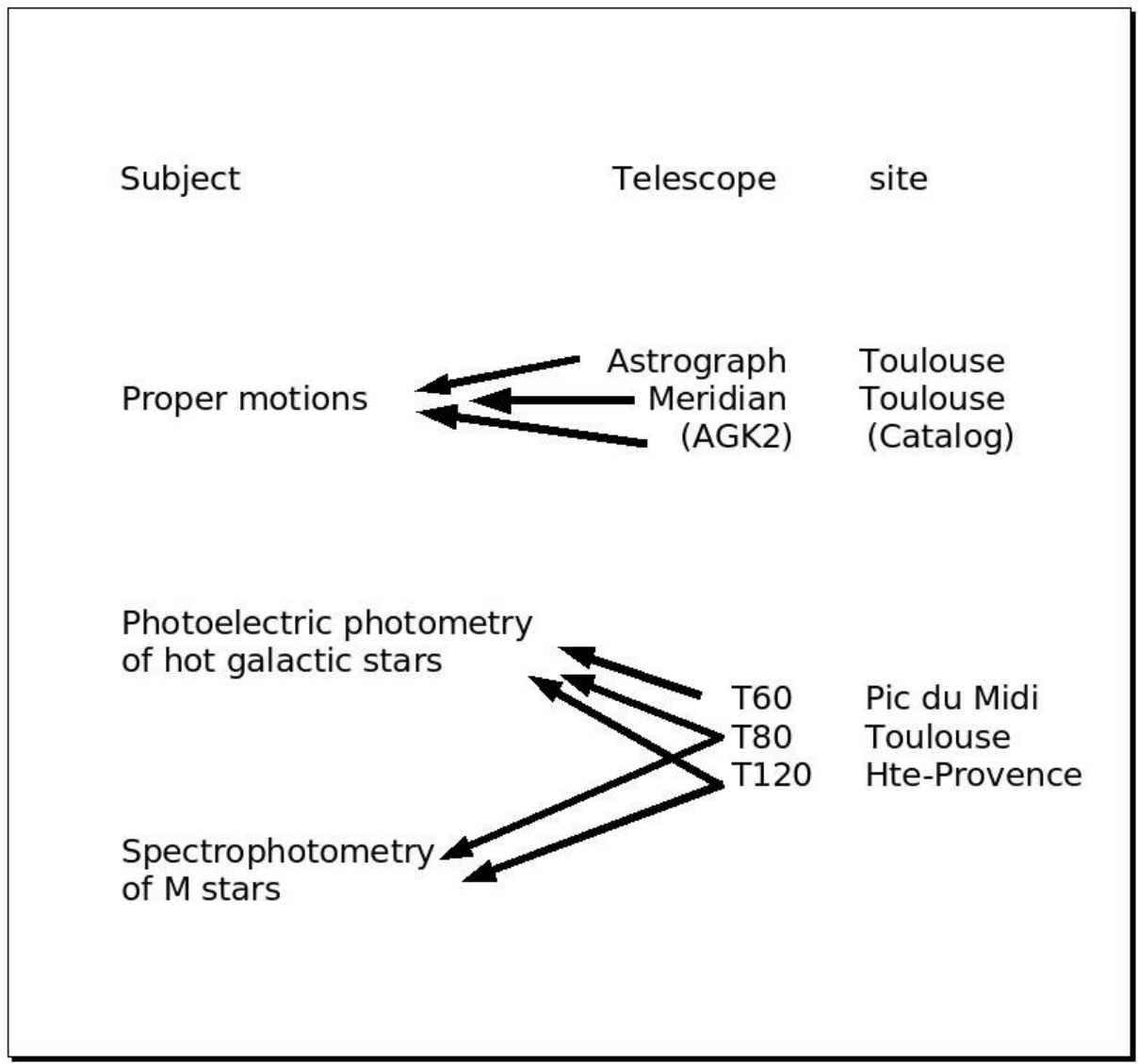

*Fig. 2 : The dedicated regime at Toulouse Observatory in 1957-58*

### III. The 80-cm telescope of Toulouse Observatory

We begin with the long and dense history of the 80-cm telescope of Toulouse Observatory, focusing on the role of the instrument in the order of practices and on its possible implication in scientific policies. We strive to understand how the different uses of such a technical object are related to the successive research projects.

The 80-cm telescope is in fact the final outcome of a long quest to equip Toulouse Observatory with a large instrument. The initial project, in 1845, called for a mural circle. When, in 1863, the director of the Observatory learned of Foucault's experiments with an 80-cm mirror, he reoriented his quest in favor of an 80-cm telescope. Numerous hurdles prevented the instrument from being acquired before 1877 and actually put to full use before 1887[11].

From 1887 to 1970, the 80-cm telescope fitted into two principal, distinct and successive techno-scientific regimes. Within the first regime, service was organized around the instruments. The purpose of astronomers was to make a detailed inventory of the night sky. The main concern of the director was thus to put all the instruments of his institution to good use toward that goal. In other words, the technical tools directed scientific activity. After a transitional phase in the 1930s, the second regime was organized around a discipline: astrophysics. The physical knowledge of stars, their composition and their structure dominated most of the scientific activity within Toulouse Observatory, especially in the period following World War 2. From then on, astrophysics structured the service.

We will attempt to distinguish how the 80-cm telescope was modified, fitted and used in these two distinct scientific cultures, examining again how it fit into a network of scientists, technicians and unanimated actors around it.

### *a) First regime : the telescope directs scientific activity (1887-1935)*

During this first period, the 80-cm telescope was an undifferentiated instrument in the global strategy of Toulouse Observatory, which was centred on astrometry. In practice, this meant spotting stars and noting their positions, as well as those of planets and their satellites. However, each instrument had a specific role which took into account its particular capacities as far as possible.

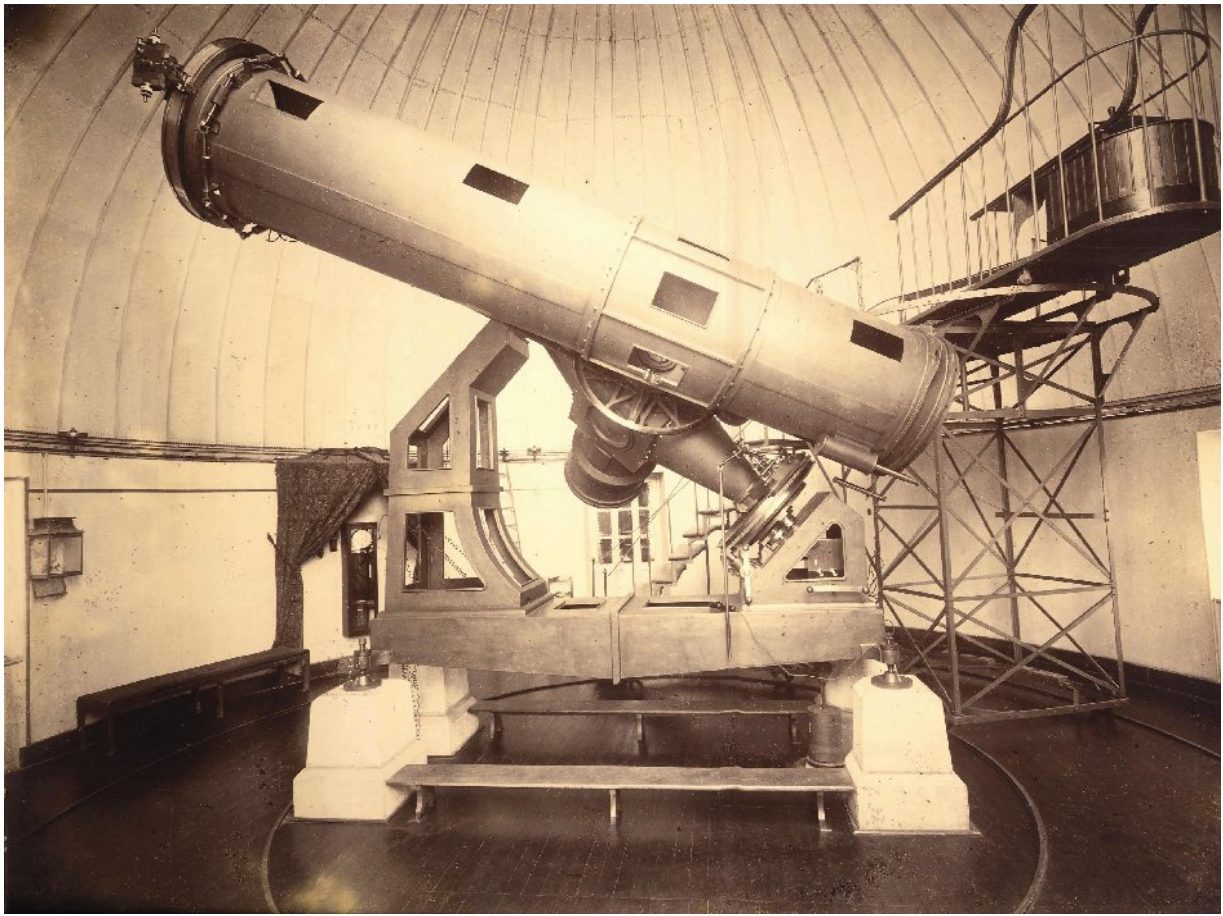

*Fig. 3 : The Toulouse Observatory 80-cm refractor (OMP Archives)*

The observatory's services were organized around each instrument and one astronomer was named responsible of it. Three astronomers worked at Toulouse Observatory in 1884, five in 1890 and seven in 1900. These astronomers were helped in the dome by one or several assistants, the caretaker in some instances. Between 1880 and 1931, the observers running the 80-cm telescope's service were, successively:

Benjamin Baillaud, Charles Fabre, Henri Andoyer, Eugène Cosserat, Henry Bourget, Alphonse Blondel and Emile Paloque.

Toulouse Observatory was one of the eighteen institutes participating in the Carte du Ciel project. The astrograph was the principal technical tool of this photographic survey, but the 80-cm telescope was also used in this context, along with all the other instruments at the site. The astronomer Henry Bourget made several "pictures of photographic calibrations for the pictures for the international catalogue".[12] In 1902, he noted that "one easily obtains on the same plate 60 stars belonging to 25 or 30 different pictures of the catalogue".[13] In 1888, the telescope was also used to "measure the numerous very weak stars in the G. Herschell Catalogue".[14]

Benjamin Baillaud's field of scientific interest was celestial mechanics. He examined Saturn's satellites with the 80-cm telescope, whose collecting power was an asset in this endeavour. The purpose there was to construct and perfect these satellites' ephemerides. The director of Toulouse Observatory indicated that "this work for which this instrument is appropriate will be pursued for a long time in view of determining elements of the orbits".[15] Baillaud's successors at the head of the establishment after 1908, Eugène Cosserat and Emile Paloque, occasionally pursued this research. Cosserat mentioned in 1910 that he had spent twenty-two evenings of the year at the telescope for "the continuation of observation of satellites of large planets [...]".[16] As for Paloque, he explained in 1926 that he had made eighteen "visual observations"[17] of Rhea, Dione, Titan, Thetis and Hyperion, as well as "32 observations of Jupiter's satellite I; 34 observations of Jupiter's satellite II; 38 observations of Jupiter's satellite III; 32 observations of Jupiter's satellite IV ".[18]

The telescope was also used under special circumstances. In 1892, the Toulouse astronomers examined the Wolf, Denning, Swift, Winnecke, Brooks I, Brooks II and Holmes[19] comets with this instrument.

This "classical" use of the telescope, centred on astrometry and on survey work, does not necessarily mean that the large collecting power of the instrument was wasted. Photographic work on the star clusters and nebulae of the New General Catalogue, begun in the 1890s[20], is visible evidence of the desire to make the most of its technical and visual possibilities.

Bourget, who was in charge of the telescope at the time, first concentrated his efforts on developing photographic techniques, making all sorts of attempts and and trying all kinds of practical combinations. In 1895, the telescope received "tiny additions intended to ease its use in work of celestial photography".[21] However, the bending of the tube caused by the guiding telescope disturbed photographic operations.[22] The year 1898 was decisive because, according to Bourget, it was the year that saw "the question of photography beyond its trial period and definitely solved".[23] Further technical improvements were made the following years : three combined pieces were introduced that, altogether, allowed the stopping at a distance of clockwork movement, the hour angle locking and unlocking from a given point of the room and the moving of the instrument in hour angle of an exact number of 2-minute lapses of time.[24] Taking pictures with repeated short exposure times was simplified by the building of an automatic shutter run by a metronome, thus ensuring a constant exposure time.

Bourget was trying to innovate in an extremely competitive scientific sector. He confessed that he had "never dreamt of rivalling the clever observers who [...] have obtained such fine images of nebulae and clusters".[25] His purpose was "completely different". He felt "that a good use of the telescope would be to try to obtain the best possible small images, appropriate to precise micrometric measurements".[26] Forced to do without the guiding telescope, Bourget suggested following "the guide star, with the help of the slow-motion levers, behind the sensitive plate through a hole made in the gelatin".[27] The Toulouse astronomers judged this solution "satisfactory"[28], because "the loss of one star on the image is greatly compensated by the improvement of the images".[29]

Henry Bourget began a wide programme of photographing stellar nebulae and clusters.[30] These images "were made with the purpose of measuring the positions of the stars they contain".[31] The astronomer therefore used "a micrometre […] on a microscope with a movable plate […] placed at his disposal by the Faculty of Sciences"[32] of Toulouse. The experimental setup designed for the Carte du Ciel project inspired Bourget, who realized that it would be "very interesting and hardly inconvenient to print on the images a grating analogous to those of the sky survey".[33] The Toulouse observer "imagined a procedure allowing the photographic printing of the grating, in bright red, on an already enveloped image".[34] This scientific undertaking begun by Bourget was pursued sporadically after his departure for Marseille Observatory in 1907. Eugène Cosserat explained in 1909 that he had "used the telescope to obtain images of clusters NGC 1960, NGC 2099, NGC 5846, NGC 6093 […]".[35] In 1926, Emile Paloque mentioned that he had "taken up the images of clusters and nebulae already photographed by H. Bourget in 1898 and 1899".[36]

To emphasize how subtle the distinction between the two regimes of usage of a telescope can be, we point out that, if Benjamin Baillaud had based his analytical perturbation theory of the minor planet Pallas on observations made at one of the telescopes rather than on archival data, this would have been a case of a telescope being used in the two regimes simultaneously.

### b) An era of transition: Paul Lacroute, the 80-cm telescope and the genesis of astrophysics in Toulouse (1935-1945)

One astronomer played a considerable role in the genesis of a genuine astrophysical project at Toulouse Observatory – this was Paul Lacroute. Already in 1934, the director Emile Paloque wished "greatly that a future nomination brings the Observatory an astronomer/physicist who is needed to get the most from this [80-cm telescope]".[37] Paul Lacroute, *agrégé* in physics and a doctor in sciences, was named "trainee assistant-astronomer starting on February 1, 1935".[38] Paloque entrusted the "great Gautier telescope"[39] to him.

Lacroute decided to use the telescope for astrophysical observations. To do this, he renewed the technical equipment associated with the telescope, which stopped being strictly for photography. From then on, auxiliary astrophysical equipment was adapted to the telescope, and the technical chain of data analysis was expanded by the acquisition of measuring instruments.

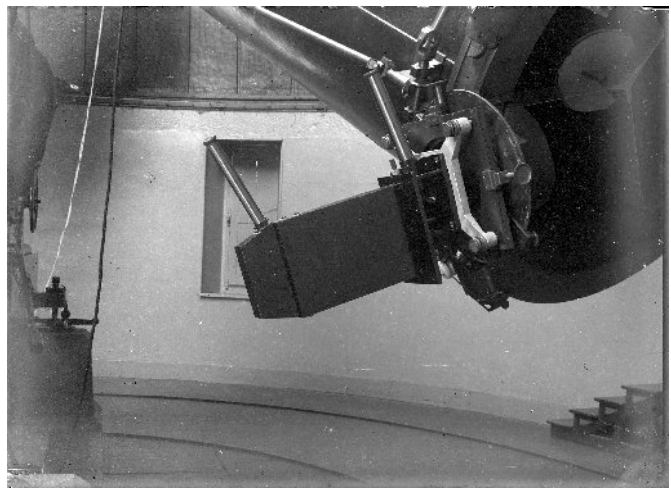

*Fig. 4: Lacroute's spectrograph (OMP Archives)*

A "radial-velocity spectrograph"[40] was ordered from Strasbourg Observatory in 1936 and delivered the following year. In 1937 and 1938, Paul Lacroute obtained "163 images of stellar spectra, of which about a hundred images were long exposures, associated most particularly with the study of hot stars with variable emission lines".[41] The variability was evidence of transient phenomena in the atmosphere of such stars. Furthermore, Lacroute himself drew up "plans for a high-dispersion spectrograph".[42] Finished in November 1938, the instrument was "immediately mounted on the telescope".[43] With this new focal instrument, Paul Lacroute continued "the study of particularly interesting irregular variables".[44]

Measuring the precise position of the centre of the lines in these spectra presupposed the use of a "recording micrometer".[45] The Caisse Nationale de la Recherche Scientifique, a new institution created in 1934[46] by Jean Perrin, provided a subsidy for the purchase of "a recording microphotometer […] from the English company 'Casella'"[47] in 1939. The technical adjustments also aimed for easier comparison of the spectra with the reference spectra, which included lines of know wavelength used to determine the others. During the years 1940-1941, the astronomer developed an assembly to juxtapose the two spectra. He "recut and polished himself the small prisms with sharp edges that allowed a better juxtaposition of the stellar spectra and the comparison spectra on the plate".[48]

The astrophysical research programme carried out by Lacroute led him to an important discovery in 1942. Helped by a Dutch astronomer, Willem Dirks, a refugee in France during the war, the Toulouse astronomer noticed "that the spectrum of star 67 Ophiuchi, presented P Cygni type emission lines".[49] The P Cygni profiles proved the existence of a gas shell in expansion around the star.

Lacroute's work and the scientific direction he moved towards transformed astronomical practice not only in the 80-cm telescope's dome, but in the very organization of the observatory. In 1942, the director of Toulouse Observatory, Emile Paloque, asserted that the purpose of "the efforts made by Mr Lacroute with remarkable activity and rare competence" were intended to make a complete organization of a "spectrographic service".[50] It was in fact an astrophysical service, to which the instrument had become subordinated, but Paloque retained the instrumentalist's mindset peculiar to the first techno-scientific level. It is therefore not surprising that he had not initiated the new regime. With Lacroute, the instrument was used for specific and innovative observations which were no longer integrated into an overall organization. Its scientific use gradually became autonomous with respect to the other instruments.

Passage from practices centred on astrometry to the deployment of the new astrophysical discipline was accelerated in Toulouse by the impossibility of exploiting the results previously obtained at the 80-cm telescope. In 1943, Lacroute obtained several photographic plates of clusters in order to "study the influence of centering on the accuracy of star positions measured on these images".[51] His purpose was to find out whether there was "any point in deducing stellar motion from the comparison of new images taken at the Telescope with old images of clusters"[52] taken by Bourget. Lacroute noted that "the result of this study was clearly negative, the slightest defect in centering causing prohibitive errors in the measured positions".[53] Because of centering defects, Bourget's astrometric work was thus unusable as first-epoch plates for measuring proper motions. The final attempt to use the 80-cm telescope in a classic astrometric undertaking was a failure and confirmed the techno-scientific change begun by Lacroute.

### c) Second regime : the telescope in the service of a scientific project (1945-1970)

Lacroute left Toulouse Observatory in 1945 to join the observatory in Strasbourg.[54] His efforts to organize astrophysical activity around the 80-cm telescope were pursued and increased by Roger Bouigue, who replaced him in 1947.

Under the impetus of Bouigue, the auxiliary instrumentation and the mechanisms for gathering information developed considerably, thus expanding the "metrological chain".[55] The 1950s were especially fruitful for the development of spectrographs adapted to the telescope. In 1952, a recording system "of the comparison spectrum was entirely reconstructed".[56] Thus, the spectrum of iron was obtained differently. Henceforth, "fluorescent tubes"[57] were used to produce other comparison spectra. It was possible to reach stars of magnitude 7, "in particular those whose spectrum presented wide atomic lines".[58] In 1955, Bouigue drew up plans and calculated the optics of a "spectrograph with prisms capable of being associated with the 80-cm telescope […] which should allow the study of weak stars in a fairly wide spectral zone (4000 to 8000A)."[59] By the following year, several nights were "dedicated to obtaining the spectra of cold M-type stars"[60] with the new spectrograph. At the same time, astronomers obtained "a Soleillet-type sampling spectrograph covering the entire spectral zone of 3600-8000 Angstroms with a perfectly flat plane".[61] The end of the decade was particularly important for the astrophysical equipment on the 80-cm telescope. During the 1959-1960 academic year, Bouigue started the construction of an electronic spectrocomparator as well as a double-grating spectrograph with double dispersion for variable stars. The former "should allow the rapid and precise measurement of stellar spectra accompanied by a comparison spectrum in view of determining the stars' radial velocity".[62] The second was intended for the "systematic study of spectra of cold variable stars".[63] The objective was "to specify the evolution of atmospheric characteristics in the course of these stars' pulsations".[64]

Roger Bouigue also innovated in the development of photoelectric photometers. In 1952 and 1953 he prepared a "cell with a Lallemand electron photomultiplier which, associated with coloured filters, should permit the determination of intensity of luminosity of bands in much more advantageous conditions than the spectrograph".[65] Installation of this new apparatus required the creation "from scratch of a photometer adapted to the focus of the large telescope in view of photometric measurements of stars and nebulae in seven different spectral bandwiths".[66] In 1954, a Meci electronic recorder was associated with the Lallemand cell in order to obtain "photoelectric measurements of very luminous stars […] in various spectral bandwidths".[67]

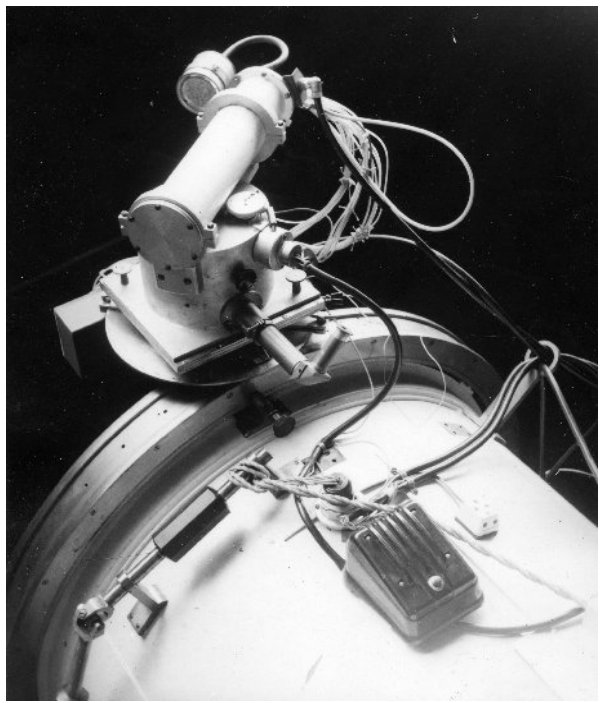

*Fig. 5 : The Lallemand electron photomultiplier (OMP Archives)*

In the same period, the telescope itself underwent only two transformations: the addition of a bonnette at the focus in 1956[68], and the optical change from Newton to false Cassegrain.[69] These modifications were not required by scientific projects but justified by a desire for greater comfort for the observer. The longer focal length meant that the astronomer was no longer required to observe from the top of a high ladder, and the bonnette simplified the mounting of auxiliary instrumentation at the focus. In fact, the telescope was no longer the object of innovation with the purpose of scientific optimization. It had reached its technical maturity and was used to the maximum of its intrinsic capability.

The addition of the auxiliary instrument even further extended the network in which the telescope was included. Maintaining these instruments required the recruitment of specific technical personnel. Four positions for technicians and assistants were created for the astrophysical service or were transferred from the Carte du Ciel service between 1950 and 1962. Furthermore, two scientists were hired in 1956 and 1957 to make observations and to carry out the new research programme.

The setting up of an astrophysical service required the creation of a network of actors and auxiliary instruments that constituted a certain number of intermediaries between the observation and the scientific result. The telescope was integrated into a techno-scientific network that aimed at imposing astrophysical practice as the heart of scientific activity at Toulouse Observatory.

The programmes of astrophysical research at Toulouse Observatory expanded. The instrument was used in collaborative studies with other observatories. Thus, starting in 1954[70], along with the establishments in Marseille, Pic du Midi and Haute-Provence, Toulouse participated in "photoelectric measurements of photographic and visual magnitudes of galactic stars".[71] Similarly, in the years 1958-1959, the observatory in Toulouse with its 80-cm telescope took part in a "campaign to examine the star β lyrae organized by Stockholm Observatory".[72] Research projects were from now on part of national and international collaboration and exchange.

This greater exchange of information and scientific data spurred the growth in Toulouse of a culture of technical exchange around the 80-cm telescope. The scientific instrument and its auxiliary equipment were gradually inserted into a national network of instrumental means. Astronomers noted in 1952 that "the very satisfying results obtained with this instrument give Toulouse Observatory important possibilities that are currently unique in France, which has attracted several Parisian researchers looking for spectra ".[73]

In 1956-1957, an astronomer at Milan-Merate Observatory, Pietro Broglia, came to Toulouse to understand "the methods of photoelectric observations", as well as "the technique of manufacturing interference filters".[74] Researchers spread the competence acquired with the telescope. They also went to other institutes to gather photometric and spectroscopic data needed for their research. Exchanges were particularly frequent with Haute-Provence Observatory.[75]
When Bouigue became director of the Toulouse institute in 1961,[76] the astrophysical service became a priority; from then on it was placed at the top in the presentation of the annual report on activity.[77] The 80-cm telescope stopped being used in the early 1970s because of light pollution.

In summary, the history of the 80-cm telescope reveals that this technical object went through two distinct regimes of usage in the course of its lifetime. It was first a general-purpose instrument used in a wide range of exploratory and/or inventory projects, which not often made full use of its technical potential. After a latency period linked to the absence of motivated users, it became the main tool of a long range research project, until environmental causes lead to its demise.

The question that now arises is whether this history, and the pattern of use that it reveals, are specific to this instrument, and thus only of anecdotical interest, or if, on the contrary, it is but one example of a general pattern for the evolving role of telescopes in astronomical research.

 **IV. Comparison with the history of the 80-cm telescope of Marseille Observatory**

It is apt to start our critical assessment of the above pattern by confronting it with the one drawn from the history of an identical telescope used in a rather similar context, that of another French provincial observatory.

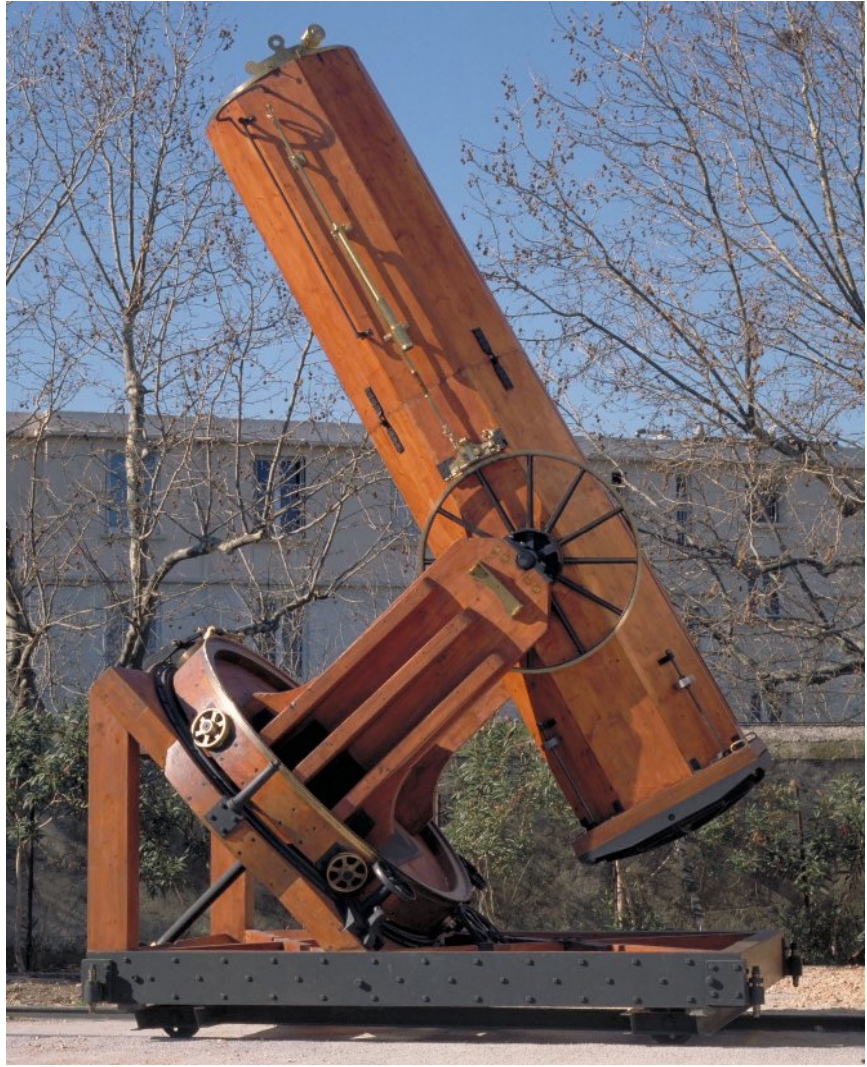

*Fig. 6 : The Marseille Observatory 80-cm refractor(OAMP Archives)*

While the history of Marseille Observatory may be rather different from that of Toulouse Observatory, the two institutes were in fact on equal footing in terms of funding, staff and instruments in the period of interest. The 80-cm telescope of Marseille Observatory went into operation in 1864, more than 20 years earlier than the one in Toulouse, and was finally dismounted in 1965, a few years before the Toulouse telescope stopped being used. The two institutes had comparable numbers of scientific staff : between three and five astronomers in the concerned period.

The initial use of the Marseille telescope is similar to that of Toulouse. Indeed, the director of the Marseille establishment, Edouard Stephan, named by Le Verrier in 1866, was a former student of Ecole Normale Supérieure, like Tisserand and Baillaud.

He began a programme to inventory nebulae[78]. The telescope was also used for observing comets, occultations of stars, transits of Mercury. These programmes are a characteristic of the general-purpose or exploratory regime.

Significant changes happened in Marseille under the influence of three academics from the Faculty of Sciences, Charles Fabry, Alfred Pérot and Henri Buisson, who dedicated themselves to developing astrophysics. They developed an interference etalon that was to be called the Fabry-Pérot etalon. It was used to measure radial velocities. In 1902, they applied their procedure to the Sun. It was not necessary to use a telescope since a simple heliostat was enough to capture solar light.[79] In 1911, they turned their attention to the Orion nebula using the 26-cm equatorial. They noted, "We hope to be able to employ it with instruments which are more powerful and better adapted to the purpose, in particular with a reflecting telescope".[80] Finally, in 1914, in collaboration with Bourget, the physicists used the 80-cm telescope to obtain 14 images with 1- to 2-hour exposure times.[81] The question arises of how the Marseille observers managed to obtain such long exposure times, since Benjamin Baillaud in Toulouse claimed it was impossible. The difference lay in the use of interferogrammes for which tracking defects have less effect on the scientific use of the image.

The Marseille scientists thus were twenty years ahead of their colleagues in Toulouse, and had passed into a different level of practice where the instrument was integrated and subordinated to a specific scientific project. Another remarkable innovation is that they published the results of their research in English in four issues of the *Astrophysical Journal*, thus reaching an international audience, which the Toulouse astronomers never did, with one notable exception (Lacroute in 1942). However, this research did not last. The three physicists were in fact not connected with the observatory, making it difficult to institutionalize a service entirely dedicated to astrophysics. The outbreak of World War I prevented the development of large-scale scientific initiatives in Marseille as well as in Toulouse.

After the war, the Marseille astronomers no longer used their telescope which was in poor condition.[82] This is rather surprising, since one of the directors of this period is Henry Bourget, who had previously put the Toulouse instrument to good use with his photographic inventory of nebulae.

The Marseille telescope was once again used when Robert Jonckheere joined the Observatory staff at the beginning of World War II. An experienced observer of binary stars since his youth (he was born in 1888), he resumed observing these systems whose relative positions he measured systematically.[83] His publications only concern the measurements themselves, with no astrophysical applications, which the astronomer leaves to future generations. In other words, this use of the telescope falls in the survey mode.

The history of the 80-cm telescope in Marseille is rather different from that of Toulouse. We do identify the two regimes – general-purpose and dedicated, but, in the present case, they appear in a cyclical order. The first regime, from 1866 to 1907, is essentially one of general-purpose, searching for and cataloging new nebulae, as well as observing targets of opportunity. The telescope then finds itself in the dedicated regime quite early on (1914) compared to the one in Toulouse, thanks to a very innovative project of Fabry and Buisson. After a long period of inactivity (1914-1941), the telescope is used by Jonckheere for a survey project, in the spirit of the general-purpose regime.

This return to scientific practice centered on the instrument and astrometry may be partially explained by Jonchkeere's training. Unlike Baillaud, Lacroute and Stephan, who were graduates of Ecole Normale Supérieure, the Marseille Observatory astronomer was self-taught, with no degrees or scientific training, and was more inclined to pursue themes of interest to amateur astronomers, thus using the telescope in the framework of the first scientific regime that made little difference in the use of technical instruments. Another, more compelling, reason is that, after WW II, astronomers preferred using the more powerful telescopes of the nearby and recently founded Haute-Provence Observatory.

We now examine historical accounts of other large telescopes to put the above results in the largest possible context.

## V. The case of other telescopes of the 20th century

### a) The Washburn Observatory 15-inch (40-cm) refractor

The succession of regimes for this telescope is identical to the one for the Toulouse instrument. In the early days of this observatory, namely between 1884 and 1922, the *Publications of Washburn Observatory* essentially report survey work with the refractor (and the meridian instrument) : micrometric observations of faint stars near bright stars of known proper motion and of double stars, observations of long-period variable stars, of the minor planet Eros in 1900-01[84]. These programmes are characteristics of the first, general-purpose, regime.

The nomination of Joel Stebbins as director in 1922 changed all this. Like Lacroute who brought innovative projects and a spectrograph to Toulouse, he contributed to Washburn Observatory a new scientific project and a new auxilliary instrument, which he mounted on the telescope, converting the latter from "a visual instrument to a dedicated photoelectric photometer"[85]. His first task was to test and improve the photometer, and then to test it on known spectroscopic binary stars for small light variations. He then developed a project for monitoring bright variable (mainly eclipsing) stars.

After this transition period, in 1930 to be precise, Stebbins embarked with two colleagues, Huffer and Whitford, on a project to investigate the reddening of stars, star clusters and galaxies, using the 15-inch refractor, but also other instruments, the Mount-Wilson 60- and 100-inch telescopes[86] - just like the Toulouse astronomers also used the Haute-Provence and Pic du Midi telescopes for their project. The result was the law of interstellar reddening. In this dedicated regime, the 15-inch was only one among several tools in a global strategy for pursuing a scientific project.

### b) The 72-inch (183-cm) telescope of the Dominion Astrophysical Observatory

This telescope presents an interesting intermediate case in our binary classification, in that the two regimes are simultaneously present after 1927.

The 72-inch telescope became operational in 1918, and was essentially used over the years for taking stellar spectra.[87] The first scientific project involved the spectroscopic observation of binary stars, mostly of O-type, for determining stellar masses (J. S. Plaskett).[88] It continued uninterrupted at least until the 1980s (with W. E. Harper and later Alan Batten), although completed by observations from the 48-inch telescope after it went into operation in 1962. This survey work fits into the first regime, despite the fact that another telescope was used, because the 72-inch could very well have been used. In 1927, Plaskett and Pearce initiated a survey of radial velocities of O- and B-stars for determining the solar motion and the constants of galactic rotation.[89] Other projects of the 1920s and 30s involved the physics of emission lines in early-type stars (H. H. Plaskett, R. K. Young).[90] While not of survey type, theses studies still fit into the first regime, since they made use of the available instrument with no alterations, thus letting the instrument lead research.

The arrival of C. S. Beals in 1927 marks the beginning of the second regime, but not the end of the first one. He implemented important changes to the existing auxiliary instrument, increasing the dispersion of the spectrograph and devising a method for including spectrophotometric calibration spots on the photographic plates, in order to study the intensity and shape of spectral lines.[91] In order to record the data from the plates, he further developed a microphotometer in 1936 and an intensitometer in 1944. The science that he did with these data was perhaps not very different from that of H. H. Plaskett and Young; the difference lies in the strategy : he adapted the telescope to his project, while the others worked the other way around. After the departure of Beals in 1946, the work in this regime was pursued by Andrew McKellar and Kenneth Wright.[92]

Further initiatives in the spirit of the dedicated regime include more. adaptations of the spectrograph for new (high) dispersions in 1938, 1946 and 1955, observations at other large telescopes (the McDonald Observatory and Curtis-Schmidt telescope at Ann Arbor) and the acquisition in 1962 of a 48-inch (122-cm) telescope to be used as an experimental adjunct of the 72-inch.

One may wonder why the changes made to the telescope by Beals and his coworkers did not eliminate the first regime of usage, which continued along with the second one. A possible answer may be found in a statement by Kenneth Wright, who became director in 1966. In his view, the research scientists were expected "to select problems that are within the capabilities of the instruments at the Observatory"[93], suggesting that the telescope should direct research, while "the general policy [was] to encourage each research scientist to carry on investigations in the fields in which he is most interested"[94], leaving room for personal initiatives, and thus for another regime of usage. But, he concluded, "… there is a strong tendency to continue along the general lines that have been established over the years".[95]

### c) The 83-cm refractor of Meudon Observatory

The history of the large refractor of Meudon Observatory has been studied in detail by Dollfus[96]. This instrument provides another example of passage from general-purpose to dedicated regime.

Jules Janssen established this observatory toward the end of the 19th century to explore the new field of physical astronomy, and the refractor that he had built in 1897 was destined for a general investigation of the physical properties of celestial objects, by means of both visual and photographic observations[97]. The instrument was perfectly adapted to this task, thanks to its long focus and high optical qualities.

The first observations, in 1898 and 1899, were of planets. The astronomers considered at one point conducting a "systematic survey of planetary surfaces"[98], and organising it into a permanent monitoring service, but, at the time, Meudon Observatory did not have the resources for such a project. The photography of star clusters became an important field of investigation for the large refractor at the turn of the century.[99] Its optical qualities enabled it to resolve the central regions of clusters into stars. Along the same line, it was possible to photograph stars surrounded by nebulosities. All these experiments were meant "to validate the potentials of the refractor".[100]

Henri Deslandres who arrived in Meudon in 1897, was a spectroscopist. The director, Jules Janssen, is then already 75 years old, and we can safely assume that Deslandres is *de facto* in charge. All the instruments should be used for spectroheliography and the measurement of radial velocities. The direct photographic observations, for which the refractor was the ideal instrument, were soon discontinued.[101] The instrument was mainly used for identifying spectroscopic double stars. It was occasionally used for other purposes : spectroscopic studies of Nova Persei (1901)[102], the rotation of Uranus (1902)[103], comet Borelly (1903)[104], Jupiter (1903-04).[105] These investigations, together with those at the other instruments, clearly put the large refractor in the general-purpose regime.

After 1903, Deslandres devoted himself to spectroheliography and lost interest in the refractor, marking an important break in the use of the instrument.[106]

The large refractor was used anew to the full extent of its visual potential upon the arrival of Eugène-Michel Antoniadi, a wealthy independent astronomer. From the end of 1910 to the 1930s, he studied the planet Mars in detail, making drawings and maps of its surface. He put an end to the controversy over Schiaparelli's "canali"[107]. He also provided detailed drawings of the planet Mercury, the surface of which is notably difficult to observe, as it is so close to the Sun.[108] He also occasionally studied Jupiter and Saturn.

In 1924, Bernard Lyot, the inventor of the coronagraph, decided to apply his newly designed polarimeter to the study of polarised light reflected off planetary surfaces, and mounted it on the large refractor, thus initiating a field of research that would be pursued well into the 1980s by the Meudon planetary astronomers, albeit mostly at other telescopes.[109]

The large refractor was also used in those days for occasionally targeting other celestial objects, such as cometary nuclei or doubtful double stars, domains where its qualities were fully exploited .

After World War II, Paul Muller, an astronomer arriving from Strasbourg Observatory, temporarily modified the regime of usage of the large refractor, in order to pursue his lifelong work on visual double stars. Between 1956 and 1974, under the sponsorship of the IAU, he secured 1000 positions and angular distances of such stars.[110] He was later transferred to Nice Observatory where he continued his quest at the 76-cm refractor of that institute.  Just like Jonckheere at Marseille Observatory, he limits his publications to measurements and the determination of orbits, putting the telescope in the survey regime of usage.

The large refractor returned to the dedicated regime in 1965, when, under the impetus of Jean Focas and in conjunction with similar observations at Pic du Midi Observatory, it was again used to study planetary surfaces.[111] The highlight of that period is probably the analysis of the martian atmosphere by Shiro Ebisawa, between 1973 and 1989. Drawings and photometric measurements obtained at the large refractor enabled him to study the seasons as well as clouds and dust on the red planet.[112]

If one sets aside the post-World War II relapse into the general-purpose regime, very much in phase with a similar pattern for the 83-cm reflector of Marseille Observatory, the large refractor of Meudon Observatory displays the now-familiar pattern of passage from a general-purpose regime of observations for their own sake, to the dedicated regime of exploration of planetary surfaces, where the instrument is perfectly adapted to the goal - imaging - and progressively becomes one element in a multi-telescope strategy for acquiring the necessary data for a single coherent project of planetary astronomy.

## VI. Conclusion

We have analysed the history over almost a century of the 80-cm reflector of Toulouse Observatory, and revealed its changing role in the conduct of astronomical research at that institute.  In the first half-century of its existence, this was a general-purpose telescope, used in a multi-faceted exploration of the night-sky and of the Sun. The telescope was then leading research, as the duty of astronomers was to make the best use of it, collecting data for future - but at that stage mostly undefined - research. The arrival of Pierre Lacroute, an astrophysicist, changed this role in the nineteen-thirties, and the telescope became dedicated to the study of stellar  spectra for astrophysical purposes, until its demise in the early nineteen-seventies. In this second regime of usage, the telescope was only one of several tools in a strategy to pursue an astrophysical project.

In order to look for a common pattern in the use of telescopes, we then briefly examined the history of similar instruments in other observatories over the same time span. That of the 80-cm telescope of Marseille Observatory was a good start, since this instrument was identical and used in a very similar scientific context, that of a provincial French observatory. This reflector was used in the general-purpose regime for most of its lifetime. Only in 1914 did it make a brief incursion into the dedicated-purpose regime, when Fabry and his collaborators used it to test and exploit their now-famous interferometer.

The two regimes of usage can be identified in the history of other telescopes, generally moving from general-purpose to dedicated regime, with occasionally a relapse into the former. For the 15-inch refractor of Washburn Observatory, the change to a dedicated project, the photometry of stars and the law of interstellar reddening, occurred in 1922 with the arrival of Joel Stebbins. At the 72-inch telescope of DAO, it happened in 1927 with the arrival of C. S. Beals, and in 1910 at the 83-cm refractor of Meudon Observatory, with Eugène Antoniadi.

The common point of all these regime changes is the arrival of a new astronomer on the staff. Only in two cases does the newcomer provoke a relapse into the general-purpose regime. In one case, that of the DAO 72-inch, the two regimes continue alongside, presumably because old habits die hard.

The two regimes of telescope usage reflect the way scientists progress, first by exploring the field, gathering data, classifying them, and only later by pursuing specific leads suggested by the patterns emerging from these data. Our proposed conceptual framework for analysing the history of telescopes is thus relevant to the period when astronomy moved from systematic, instrument-led exploration to more focussed, project-led research.

The trend for telescopes to move from general to dedicated purpose continues to this day, when large multi-purpose telescopes such as the 4 ESO VLTs, the 2 Kecks in Hawaii, are used to explore the cosmic frontier in all fields, while other telescopes, such as the SLOAN 2.5m telescope at Apache Point Observatory, ESA's Hipparcos space astrometry mission, or the Wilkinson Microwave Anisotropy Probe of NASA, have been designed for specific tasks. However, the situation is now much more complex than in the past century, and our simple conceptual framework generally does not apply, as the various actors around large telescopes have different goals. The managers of telescopes are concerned with optimising the output of their instrument in terms of data and publications, while for science teams the telescope is but one tool in their strategy.

While recent works in the social studies of science provide numerous examples of instruments built - in part or totally - by scientists to pursue their own research[113], analyses of the structuring effect of an instrument, completely organising the research of a scientific realm are scarce[114], and the main contribution of this paper is perhaps to show that the historical analysis of a scientific instrument can combine these two approaches with profit.

We have revealed a permanent structuring tension between institutionalised science policies and the relative autonomy of astronomers in their research projects, leading to a more intense (and possibly efficient) use of the telescope. Such a tension is probably even more striking in contemporary research, as we alluded to above. The variety and multiplicity of contexts allow one to better understand how one or the other component of the tension prevails, and provide a wider view of the actors' field of action and of the possible constructive outside effects (institutions, research programs).

The use of a "long time span"[115] in the study of instruments provides a balance between the macro approach which tends to underestimate local arrangements, and the micro analysis which may neglect the wider stakes of science policies. In maintaining the interplay between macro and micro, one can grasp, for a given instrument, the importance of individual opportunities, of collective choices, the mode of integrations of technological innovations, as well as paradigm changes and other mutations in science. The role of these various elements depends on the epoch and the context, and at a given stage define the prevailing potentials and stakes.

In closing, we suggest that the concept of two regimes - general purpose or exploratory and dedicated - presented here be further tested and extended by investigating the interplay between the academic and industrial spheres over the same period. Surely the instrument makers must have influenced the path of scientific research, or was it the other way around ?

**ACKNOWLEDGMENTS**

We thank the referee, Dr. Batten, for very valuable comments which lead to major improvements in the paper.

**REFERENCES**

Baillaud, Benjamin. 'Observatoire de Toulouse' in *Rapport sur les observatoires de province*. Paris: Imprimerie Nationale, 1888 : 31-34.
Baillaud, Benjamin. 'Observatoire de Toulouse' in *Rapport sur les observatoires de province*. Paris: Imprimerie Nationale, 1890 : 32-34 .
Baillaud Benjamin 'Observatoire de Toulouse' in *Rapport sur les observatoires de province*. Paris : Imprimerie Nationale, 1892 :  42-46.
Baillaud, Benjamin. 'Observatoire de Toulouse' in *Rapport sur les observatoires de province.* Paris: Imprimerie Nationale, 1895 : 40-45.
Baillaud, Benjamin. 'Observatoire de Toulouse' in *Rapport sur les observatoires de province*. Paris: Imprimerie Nationale, 1897 : 44-51 .
Baillaud, Benjamin. 'Observatoire de Toulouse' in *Rapport sur les observatoires de province*. Paris: Imprimerie Nationale, 1899 : 49-56.
Baillaud, Benjamin. 'Observatoire de Toulouse' in *Rapport sur les observatoires de province*. Paris : Imprimerie Nationale, 1902 : 111-120.
Bell Trudy E. 'Money and Glory' *The Bent of Tau Beta Pi* 117 (2006) : 13-20.
Bennett, Jim A. 'On the power of penetrating into space: the telescopes of William Hershel' *Journal for the History of Astronomy*,19 (1976) : 75-108
Bouigue, Roger. *Rapport présenté au Conseil de l'Université par M. Bouigue, Directeur de Toulouse, sur l'état actuel de cet Etablissement et sur les travaux accomplis pendant l'année scolaire 1961-1962 et sur les perspectives de développement envisagées*. Toulouse, 1962).
Bouigue, Roger. *Rapport sur l'activité de l'Observatoire de Toulouse pendant l'année scolaire 1965-1966, présenté par M. Bouigue, Directeur, Toulouse*. Toulouse, 1966.
Bourget, Henri. *Photographie des nébuleuses et des amas stellaires*. Paris, 1900.
Braudel, Fernand.  *Écrits sur l'histoire*. Paris Flammarion, 1969.
Buisson, Henri and Fabry, Charles. 'An application of interference to the study of the Orion nebula' *Astrophysical Journal*. 40 (1914) : 241-258.
Clarke A. and Fujimura J. 'Quels outils ? Quelles tâches ? Quelle adéquation ?' in *La matérialité des sciences. Savoir-faire et instruments dans les sciences de la vie*,  edited by A. Clarke A. and J. Fujimura. Marsat : Synthélabo, 1996 : 17-68.
Cosserat, Eugène. 'Observatoire de Toulouse' in *Rapport sur les observatoires de province*. Paris, 1910 : .
Cosserat, Eugène. 'Observatoire de Toulouse' in *Rapport sur les observatoires de province*. Paris, 1909.
Davoust, Emmanuel. *L'observatoire du Pic-du-Midi. Cent ans de vie et de science en haute montagne*. Paris : éditions du CNRS.
Dollfus, Audouin. 'La grande lunette de l'Observatoire de Meudon, 1. Construction et premières observation' *L'astronomie* 120 (2006) : 82-90.
Dollfus, Audouin. 'La grande lunette de l'Observatoire de Meudon, 2' *L'astronomie* 120 (2006) : 144-155.

Dollfus, Audouin. *La grande lunette de Meudon. Les yeux de la découverte*. Paris, éditions du CNRS, 2006.
Dollfus, Audouin. *La grande lunette de Meudon. Les yeux de la découverte*. Paris : éditions du CNRS, 2006.
Fabry, Charles and Buisson, Henri. 'Application of the interference method to the study of nebulae' *Astrophysical Journal* 33 (1911) : 406-409.
Fabry, Charles and Perot A. 'Measures of absolute wave-lengths in the solar spectrum and in spectrum of iron' *Astrophysical Journal* 15 (1902) : 261-273.
Fabry, Charles, and Perot, A. 'Measures of absolute wave-lengths in the solar spectrum and in spectrum of iron' *Astrophysical Journal* 15 (1902) : 73-96.
Gibson, Brad K. 'Liquid Mirror Telescopes: History' *Journal of the Royal Astronomical Society of Canada* 85l (1985): 158-171.
Giclas, H.L. 'History of the 13-Inch Photographic Telescope and its Use Since the Discovery of Pluto' *Icarus* 44 (1980): 7-11.
Graham Smith, F. and J. Dudley. 'The Isaac Newton Telescope' *Journal for the History of Astronomy*, 13 (1982) : 1-18.
Hoskin, Mickaël,.'Herschel's 40ft Reflector: Funding and Functions', *Journal for the History of Astronomy*, 34 (2003) : 1-32.
Jonckheere, Robert. 'Etoiles doubles nouvelles découvertes à l'Observatoire de Marseille' *Journal des Observateurs* 24 (1941) : 21-25.
Jonckheere, Robert. 'Etoiles doubles nouvelles découvertes à l'Observatoire de Marseille' *Journal des Observateurs*, 24 (1941) : 69-72.
Jonckheere, Robert. 'Etoiles doubles nouvelles découvertes à l'Observatoire de Marseille' *Journal des Observateurs*, 24 (1941) : 93-96.
Lacroute, Paul and W.H. Dirks. 'Note on the spectrum of 67 Ophiuchi' *Astrophysical Journal*, 96 (1942) :.481.
Lamy Jérôme. 'The chaotic genesis of a scientific instrument : the 80-cm telescope at Toulouse observatory (1848-1877)' *Bulletin of the Scientific Instrument Society* 99 (2009) : 2-8.
Latour, Bruno. *La science en action*. Paris : Gallimard, 1989.
Leibl, David S. and Flucke, Christopher. 'Investigations of the interstellar medium at Washburn Observatory, 1930-58' *Journal of Astronomical History and Heritage*, 6 (2004) : 85-94.
Lovell, Bernard. 'The Early History of the Anglo-Australian 150-inch Telescope (AAT)' *The Quarterly Journal of the Royal Astronomical Society* 26 (1985): 393-455.
Paloque, Emile. *Rapport sur les observatoires de province*. Paris, 1926. .
Paloque, Emile. 'Observatoire de Toulouse' in *Rapport sur les observatoires de province*. Paris, 1934 : 168-178:
Paloque, Emile. *Rapport présenté au Conseil de l'Université par M. Paloque, directeur de l'Observatoire de Toulouse, sur l'état actuel de cet établissement et sur les travaux accomplis pendant l'année 1934-1935*. Toulouse, 1935.
Paloque, Emile. *Rapport présenté au Conseil de l'Université par M. Paloque, Directeur de l'Observatoire de Toulouse, sur l'état actuel de cet établissement et sur les travaux accomplis pendant l'année scolaire 1935-1936*. Toulouse, 1936.
Paloque, Emile. *Rapport présenté au Conseil de l'Université par M. Paloque, Directeur de l'Observatoire de Toulouse, sur l'état actuel de cet établissement et sur les travaux accomplis pendant l'année scolaire 1937-1938*. Toulouse, 1938.
Paloque, Emile. *Rapport présenté au Conseil de l'Université par M. Paloque, Directeur de l'Observatoire de Toulouse, sur l'état actuel de cet établissement et sur les travaux accomplis pendant l'année scolaire 1938-1939*. Toulouse, 1939.
Paloque, Emile. *Rapport présenté au Conseil de l'Université par M. Paloque, Directeur de l'Observatoire de Toulouse, sur l'état actuel de cet établissement et sur les travaux accomplis pendant l'année scolaire 1940-1941*. Toulouse, 1941.
Paloque, Emile. *Rapport présenté au Conseil de l'Université par M. Paloque, Directeur de l'Observatoire de Toulouse, sur l'état actuel de cet établissement et sur les travaux accomplis pendant l'année scolaire 1941-1942*. Toulouse, 1942.
Paloque, Emile. *Rapport présenté au Conseil de l'Université par M. Paloque, Directeur de l'Observatoire de Toulouse, sur l'état actuel de l'Etablissement et sur les travaux accomplis pendant l'année 1945-1946*. Toulouse, 1946.
Paloque, Emile. *Rapport présenté au Conseil de l'Université par M. Paloque, Directeur

de l'Observatoire de Toulouse, sur l'état actuel de l'Etablissement et sur les travaux accomplis pendant l'année 1951-1952*. Toulouse, 1952.

Paloque, Emile. *Rapport présenté au Conseil de l'Université par M. Paloque, Directeur de l'Observatoire de Toulouse sur l'état actuel de cet Etablissement et sur les travaux accomplis pendant l'année scolaire 1952-1953*. Toulouse, 1953.

Paloque, Emile. *Rapport présenté au Conseil de l'Université par M. Paloque, Directeur de l'Observatoire de Toulouse, sur l'état actuel de cet Etablissement et sur les travaux accomplis pendant l'année scolaire 1953-1954.* Toulouse, 1954.

Paloque, Emile. *Rapport présenté au Conseil de l'Université par M. Paloque, Directeur de l'Observatoire de Toulouse, sur l'état actuel de cet Etablissement et sur les travaux accomplis pendant l'année scolaire 1954-1955.* Toulouse, 1955.

Paloque, Emile. *Rapport présenté au Conseil de l'Université par M. Paloque, Directeur de l'Observatoire de Toulouse, sur l'état actuel de cet Etablissement et sur les travaux accomplis pendant l'année scolaire 1955-1956*. Toulouse, 1956.

Paloque, Emile. *Rapport présenté au Conseil de l'Université par M. Paloque, Directeur de l'Observatoire de Toulouse, sur l'état actuel de cet Etablissement et sur les travaux accomplis pendant l'année scolaire 1956-1957.* Toulouse, 1957.

Paloque, Emile. *Rapport présenté au Conseil de l'Université par M. Paloque, directeur de l'Observatoire sur l'état actuel de cet établissement et sur les travaux accomplis pendant l'année scolaire 1958-1959*. Toulouse, 1959.

Paloque, Emile. *Rapport présenté au Conseil de l'Université par M. Paloque, Directeur de l'Observatoire de Toulouse, l'état actuel de cet Etablissement et sur les travaux accomplis pendant l'année scolaire 1959-1960*. Toulouse, 1960.

Picard, Jean-François. *La république des savants. La Recherche française et le CNRS* . Paris : Flammarion, 1990.

Shinn, Terry, 'The Bellevue grand électroaimant, 1900-1940 : Birth of a research-technology community', *Historical Studies in the Physical Sciences* 24 (1993) : 157-187.

Shinn, Terry, 'The Resarch-Technology Matrix : German Origins, 1860-1900' in *Instrumentation. Between Science, State and Industry* edited by B. Joerges and T. Shinn. Dordrecht: Kluwer Academic Publishers, 2001 : 29-48.

Tobin William. 'Foucault's invention of the silvered glass reflecting telescope and the history of his 80-cm reflector at the Observatoire de Marseille' *Vistas in Astronomy* 30 (1987) : 153-184.

Véron, Philippe. 'L'équatorial de la tour de l'est de l'Observatoire de Paris' *Revue d'histoire des sciences*, 61 (2003) : 191-220.

Vinck, Dominique. *Du Laboratoire aux réseaux. Le travail scientifique en mutation*, Luxembourg, 1992.

Wright, K.O. 'Fifty years at the Dominion astrophysical observatory' *The Journal of the Royal Astronomical Society of Canada*, 62 (1968) : 269-286.

**NOTES**

[1]. Bennett, 'On the power of penetrating into space: the telescopes of William Hershel, ' Hoskin, 'Herschel's 40ft Reflector: Funding and Functions.'
[2]. Giclas, 'History of the 13-Inch Photographic Telescope and its Use Since the Discovery of Pluto.'
[3]. Gibson, 'Liquid Mirror Telescopes: History.'
[4]. Graham Smith, andDudley , 'The Isaac Newton Telescope.'
[5]. Lovell, 'The Early History of the Anglo-Australian 150-inch Telescope (AAT).',
[6] . Bell, 'Money and Glory.'
[7]. Véron, 'L'équatorial de la tour de l'est de l'Observatoire de Paris.'
[8]. Davoust, *L'observatoire du Pic-du-Midi,* 129-150.
[9]. Dollfus, *La grande lunette de Meudon.* See also: Dollfus, 'La grande lunette de l'Observatoire de Meudon, 1. Construction et premières observations,' Dollfus, 'La grande lunette de l'Observatoire de Meudon, 2.'
[10]. Tobin, 'Foucault's invention of the silvered glass reflecting telescope and the history of his 80-cm reflector at the Observatoire de Marseille.'
[11]. Lamy, 'The chaotic genesis of a scientific instrument : the 80-cm telescope at Toulouse observatory (1848-1877).'
[12]. Baillaud, 'Observatoire de Toulouse', 1902, 53.
[13] . *Ibid.*, 53.
[14]. Baillaud, 'Observatoire de Toulouse', 1888, 37.
[15]. Baillaud, 'Observatoire de Toulouse', 1890, 40.
[16]. Cosserat, 'Observatoire de Toulouse', 1910, 62.
[17]. Paloque, *Rapport sur les observatories de province*, 64.
[18] . *Ibid.*, 65.
[19]. Baillaud, 'Observatoire de Toulouse', 1892, 40.
[20] . Bourget, *Photographie des nébuleuses et des amas stellaires*, 1.
[21]. Baillaud, 'Observatoire de Toulouse', 1895) 38-39.
[22]. Baillaud, 'Observatoire de Toulouse', 1897, 38-39.
[23] . Henri Bourget, Report of the year 1898, 2R 112, Municipal Archive of Toulouse (MAT)
[24]. Baillaud, 'Observatoire de Toulouse', 1899) 53.
[25] . Bourget, *Photographie des nébuleuses et des amas stellaires*, 1.
[26] . *Ibid.*, 1.
[27] . *Ibid.*, 3.
[28]. Baillaud, 'Observatoire de Toulouse', 1897, 41.
[29] . *Ibid.*, 42.
[30]. Bourget *Photographie des nébuleuses et des amas stellaires*, 1.
[31]. Baillaud, 'Observatoire de Toulouse', 1899, 55.
[32]. *Ibid.*, 55.
[33]. *Ibid.*, 55.
[34]. *Ibid.*, 55.
[35]. Cosserat, 'Observatoire de Toulouse' 1909, 67.
[36]. Paloque, *Rapport sur les observatories de province*, 1926, 64.
[37]. Paloque, 'Observatoire de Toulouse', 1934, 73.
[38] . Paloque, *Rapport présenté au Conseil de l'Université*, 1935, 168.
[39]. *Ibid.*, 171.
[40]. Paloque, *Rapport présenté au Conseil de l'Université*, 1936, 179.
[41]. Paloque, *Rapport présenté au Conseil de l'Université* , 1938, 171.
[42]. Paloque, *Rapport présenté au Conseil de l'Université*, 1936, 179.
[43]. Paloque, *Rapport présenté au Conseil de l'Université*, 1939, 137.
[44]. *Ibid*. 137.
[45]. *Ibid*. 137.
[46] . Picard, *La république des savants*, 49.
[47]. Paloque, *Rapport présenté au Conseil de l'Université*, 1939, 137.
[48]. Paloque, *Rapport présenté au Conseil de l'Université*, 1941, 146.
[49]. Paloque, *Rapport présenté au Conseil de l'Université*, 1942, 176. Lacroute, Dirks, 'Note on the spectrum of 67 Ophiuchi', 481.
[50]. Paloque, *Rapport présenté au Conseil de l'Université*, 1942), 175.
[51]. *Ibid.,* 130.

52. *Ibid.*, 130.
53. *Ibid.*, 130.
54. Paloque, *Rapport présenté au Conseil de l'Université*, 1946, 149.
55. Latour, *La science en action.*, 606.
56. Paloque, *Rapport présenté au Conseil de l'Université*, 1952, 179.
57. *Ibid.*, 179.
58. *Ibid.*, 180.
59. Paloque, *Rapport présenté au Conseil de l'Université*, 1956, 5.
60. Paloque, *Rapport présenté au Conseil de l'Université*, 1957, 224.
61. *Ibid.* , 225.
62. Paloque, *Rapport présenté au Conseil de l'Université* , 1960, 344.
63. *Ibid.*, 344.
64. *Ibid.*, 344.
65. Paloque, *Rapport présenté au Conseil de l'Université*, 1953, 157.
66. Paloque, *Rapport présenté au Conseil de l'Université*, 1954, 189.
67. Paloque, *Rapport présenté au Conseil de l'Université*, 1955, 195.
68. Paloque, *Rapport présenté au Conseil de l'Université*, 1957, 225.
69. Bouigue, *Rapport sur l'activité de l'Observatoire de Toulouse*, 1966, 680.
70. Paloque, *Rapport présenté au Conseil de l'Université*, 1955, 195.
71. Paloque, *Rapport présenté au Conseil de l'Université*, 1960, 343.
72. Paloque, *Rapport présenté au Conseil de l'Université*, 1959, 287.
73. Paloque, *Rapport présenté au Conseil de l'Université*, 1952, 180.
74. Paloque, *Rapport présenté au Conseil de l'Université*, 1957, 220.
75. For example, Bouigue, Chapuis, Pédoussaut and Rochette made three observing runs in January, May and July 1959 with the 120-cm telescope of Haute-Provence Observatory. (Paloque, *Rapport présenté au Conseil de l'Université*, 1959, 285). In October 1959 and June 1960, Bouigue and Pédoussaut used the large spectrograph with the 193-cm telescope. (Paloque, *Rapport présenté au Conseil de l'Université*, 1960, 343).
76. Bouigue, *Rapport présenté au Conseil de l'Université*, 1962, 286.
77. *Ibid*, 287. This is the first time that the astrophysical service is placed at the head.
78. Tobin, 'Foucault's invention of the silvered glass reflecting telescope, 167.
79. *Ibid.*, 171.
80. Fabry and Perot, 'Measures of absolute wave-lengths in the solar spectrum and in spectrum of iron,' Fabry and Perot, 'Measures of absolute wave-lengths in the solar spectrum and in spectrum of iron.'
81. Fabry and Buisson, 'Application of the interference method to the study of nebulae.'
82. Buisson and Fabry, 'An application of interference to the study of the Orion nebula,' 246.
83. Jonckheere, 'Etoiles doubles nouvelles découvertes à l'Observatoire de Marseille,' Jonckheere, 'Etoiles doubles nouvelles découvertes à l'Observatoire de Marseille,' Jonckheere, 'Etoiles doubles nouvelles découvertes à l'Observatoire de Marseille.'
84. Leibl and Flucke, 'Investigations of the interstellar medium at Washburn Observatory, 1930-58,' 85-86.
85. *Ibid.*, 87.
86. *Ibid.*, 88.
87. Wright, 'Fifty years at the Dominion astrophysical observatory,' 271.
88. *Ibid.*, 269.
89. *Ibid.*, 276.
90. *Ibid.*, 275.
91. *Ibid.*, 277.
92. *Ibid.***,** 280-281.
93. *Ibid.***,** 271
94. *Ibid.*, 271.
95. *Ibid.*, 271.
96. Dollfus, *La grande lunette de Meudon.*
97. *Ibid.*, 20
98. *Ibid.*, 79
99. *Ibid.*, 79-81.
100. *Ibid.*, 83.

[101]. *Ibid.*, 87-100.
[102]. *Ibid.*, 90-94.
[103]. *Ibid.*, 96-97.
[104]. *Ibid.*, 98.
[105]. *Ibid.*, 99-100.
[106]. *Ibid.*, 98.
[107]. *Ibid.*, 100-114.
[108]. *Ibid.*, 116-120.
[109]. *Ibid.*, 127-135.
[110]. *Ibid.*, 150.
[111]. *Ibid.*, 151-153.
[112]. *Ibid.*, 165-166.
[113]. Clarke and Fujimura, 'Quels outils ? Quelles tâches ? Quelle adéquation ?'
[114]. For exemple : D. Vinck, *Du Laboratoire aux réseaux.*
[115]. Braudel, *Écrits sur l'histoire*.

Jérôme Lamy is a post-doctoral student at Toulouse University (France). He is a historian of astronomy. His astronomical research interests include the relationships between observatories and universities, scientific instruments, space research in the twentieth century and more generally the scientific institutions. He is the author of more than twenty research papers, and the book *L'Observatoire de Toulouse aux XVIII$^e$ et XIX$^e$ siècles. Archeologie d'un espace savant* (2007). He is the editor of the book *La Carte du Ciel. Histoire et actualité d'un projet scientifique international* (2008). He received, in 2004, the Sydney Forado prize for his PhD.

Emmanuel Davoust is a senior astronomer at Observatoire Midi-Pyrénées, where he conducts research on multivariate analyses of samples of galaxies and of globular clusters. He also gathers observations at large telescopes in view of their statistical analysis. He is in charge of the Heritage Commission of his institute, and thus also works in the field of history of astronomy. Recent publications : Sharina, M., Davoust, E., 'Globular cluster content and evolutionary history of NGC 147', *Astronomy and Astrophysics*, 497, 65, 2009 ; Davoust, E.; Contini, T., 'Starbursts in barred spiral galaxies. VI. HI observations and the K-band Tully-Fisher relation', *Astronomy and Astrophysics*, 416, 515, 2004 ; Davoust, E., *L'Observatoire du Pic du Midi, cent ans de vie et de science en haute montagne*", Paris, CNRS-Editions, 2000.